\documentclass[onecolumn,12pt,conference]{IEEEtran}
\usepackage{longtable}
\usepackage{amsmath}
\usepackage{cancel}
\usepackage{amssymb}
\usepackage{graphicx}
\usepackage{longtable}
\usepackage{multirow}
\usepackage{array}
\usepackage[labelsep=period]{caption}
\usepackage{setspace}
\usepackage{float}
\usepackage{subfig}
\usepackage{lettrine}
\usepackage[noadjust]{cite}
\usepackage{cleveref}
\usepackage[letterpaper, left=1in, right=1in, bottom=1in, top=0.75in]{geometry}

\IEEEoverridecommandlockouts \IEEEpubid{\makebox[\columnwidth]{ 978-1-5386-3531-5/17/\$31.00~\copyright~2017 IEEE \hfill} \hspace{\columnsep}\makebox[\columnwidth]{ }}
\begin{document}
%
\title{Error Performance of Wireless Powered Cognitive Relay Networks with Interference Alignment}

\author{$\textrm{Sultangali~Arzykulov}^{*}$,~ $\textrm{Galymzhan~Nauryzbayev}$,~$\textrm{Theodoros A. Tsiftsis}^{*}$\\
	and
	$\textrm{Mohamed~Abdallah}^{\ddagger}$\\
	\IEEEauthorblockA{$^*$School of Engineering, Nazarbayev University, Astana, Kazakhstan\\
		$^{\ddagger}$Division of Information and Computing Technology, College of Science and Engineering, \\
		Hamad Bin Khalifa University, Qatar Foundation, Doha, Qatar\\
		Email: \{sultangali.arzykulov, theodoros.tsiftsis\}@nu.edu.kz,\\
		nauryzbayevg@gmail.com, moabdallah@hbku.edu.qa}
}


%


\maketitle

\begin{abstract}
This paper studies a two-hop decode-and-forward underlay cognitive radio system with interference alignment technique. An energy-constrained relay node harvests the energy from the interference signals through a power-splitting (PS) relaying protocol. Firstly, the beamforming matrices design for the primary and secondary networks is demonstrated. Then, a bit error rate (BER) performance of the system under perfect and imperfect channel state information (CSI) scenarios for PS protocol is calculated. Finally, the impact of
the CSI mismatch parameters on the BER performance is simulated.
\end{abstract}

%
\IEEEpeerreviewmaketitle

\section{Introduction}
\IEEEPARstart{C}{ognitive radio} (CR) has attracted a significant attention by being an intelligent technology that can utilize radio spectrum and increase the spectral efficiency in wireless networks \cite{Goldsmith}. The main idea of CR is to provide secondary users (SUs), which are unlicensed nodes, with possibility to communicate in a licensed spectrum on the condition that primary users (PUs) should not receive harmful interference \cite{Goldsmith,Arzykulov}. There are three types of spectrum sharing CR paradigms, namely, interweave, underlay and overlay \cite{Goldsmith}. The interference mitigation at receivers can be managed by a promising technique named interference alignment (IA). IA provides free signaling dimensions by aligning the interference into one subspace \cite{galym1,galym2}. IA can be also applied in CR to cancel interference at primary and secondary receivers. A mitigation of the severe effect of interference at primary receivers allows secondary transmitters to increase their transmit power which consequently leads to an improvement of the performance of the secondary network (SN) \cite{Amir}. In \cite{Tang}, degrees of freedom (DoFs) of the network were increased by implementing an IA technique in a multiple-input multiple-output (MIMO) relay CR. 

Another promising technique, known as energy harvesting (EH), which harvests energy from ambient signals through time-switching (TS) and power-splitting (PS), was introduced in \cite{Nasir,GN}.  IA and simultaneous wireless information and power transfer (SWIPT) in MIMO networks were jointly studied in \cite{Zhao1}, where an improvement of the performance of the network was analyzed through a dynamical selection of users as  EH or information decoding (ID) terminals. The EH-based CR network was studied in \cite{Park}, where authors developed a spectrum sensing policy in TS mode to guarantee EH possibility for SUs from primary signals. An identical system was studied in \cite{Zheng}, where an optimal information and energy cooperation methods between primary and secondary networks was investigated. Finally, the work in \cite{Wang} represented a resource allocation method in EH-based CR network with imperfect channel state information (CSI) cases.  

In this paper, we study an underlay IA-based CR with an energy-restricted relay operating in PS mode. The performance of both the primary network (PN) and SN after interference mitigation is analyzed. In particular, a bit error rate (BER) performance of the proposed system model is calculated for PS relaying protocol under different imperfect CSI cases.

\section{System Model}
\label{sec:system model}
\begin{figure}[!h]
	\centering
	\includegraphics[width=0.4\columnwidth]{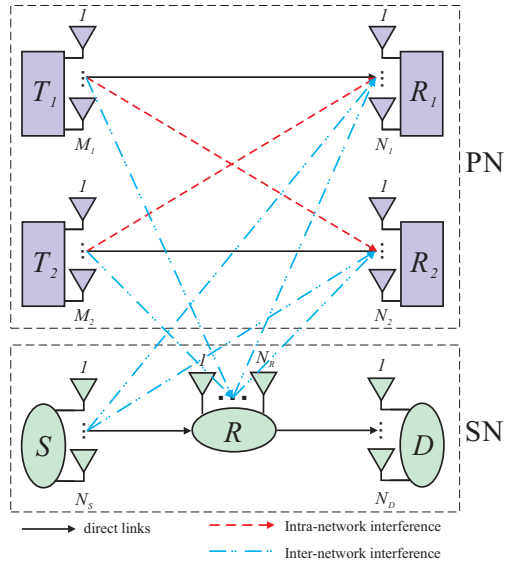}
	\caption{IA and EH-based CRN with two PUs and one SU sharing the spectrum simultaneously.}
	\label{system_model}
\end{figure}

The proposed system model is composed of a PN with two pairs of PUs and a SN with three SUs. Each primary transmitter ($T_{i}$) transmits to its corresponding receiver by interfering with another primary receiver ($R_{j}$) and relay as shown in Fig. \ref{system_model}. The SN is composed of  a source ($S$), a relay ($R$) and a destination ($D$) nodes. An energy constrained $R$ operates in decode-and-forward (DF) half-duplex mode by relaying the signal from $S$ to $D$ in two time periods. $R$ uses harvested energy from interference signals as its transmit power, while $S$ and $D$ supplied with stationary power sources. Also, it is assumed that $D$ is located far from PN and does not receive any interference. All nodes of the network are assumed to have MIMO antennas. We also assume that all interference at $R_{j}$ are canceled by IA technique, thus, $S$ and $R$ are not restricted by the level of transmit power. Another assumption is that the channels remain constant during a transmission block time $T$, but vary independently from one block to another. The definition of channel links between nodes can be denoted by the next. For channels of PN nodes,
$\mathbf{H}_{j,i}^{[k]}\in\mathbb{C}^{N_j\times M_i},~\forall i,j \in \{1,2\}$ denotes the channel between  $R_{j}$ and $T_{i}$, where superscript $k$ indicates a certain time period  when the data transmission occurs. $N_j$ and $M_i$ are the numbers of antennas at $R_{j}$ and $T_{i}$, respectively. For channels of SN nodes, $\mathbf{H}_{R,S}$ and $\mathbf{H}_{D,R}$ denote the channel links related to the $S$-$R$ and $R$-$D$ transmissions while the inter-network channels are given by $\mathbf{H}_{j,R}\in\mathbb{C}^{N_j\times N_R}$, $\mathbf{H}_{j,S}\in\mathbb{C}^{N_j\times N_S}$ and $\mathbf{H}_{R,i}\in\mathbb{C}^{N_R\times M_i}$, where $N_S$, $N_R$ and $N_D$ denote the numbers of antennas at $S$, $R$ and $D$, respectively. Each entry of any matrix $\mathbf{H}$ is assumed to be independent and identically distributed (i.i.d.) random variables according to $\mathcal{CN}(0,1)$, where $\mathcal{CN}(0,1)$ denotes the complex normal distribution with zero mean and unit variance. Also, note that each channel link is characterized by the corresponding distance and path-loss exponent denoted by $d_{m,n}$ and $\tau_{m,n},~\forall m\in\{1,2,R,D\},~\forall n\in\mathcal{A} = \{1,2,S,R\}$, respectively. 
We assume that each node is deployed with $N$ antennas ($M_i=N_j=N_{S}=N_{R}=N_{D} = N$) and IA is exploited at $R$ and $R_{j}$, accordingly. Therefore, each transmit node $l$ with the power $P_l$ employs a precoding matrix $\mathbf{V}_{l}\in\mathbb{C}^{(M_l~\textrm{or}~N_l)\times f_l}$, with  $\textrm{trace}\{\mathbf{V}_l\mathbf{V}_l^H\}=1,~\forall l \in\mathcal{A}$, where $f_l$ is the number of the transmitted data streams. Then, each receive node, except $D$, employs the interference suppression matrix $\mathbf{U}_t \in\mathbb{C}^{N_t\times f_t},~\forall t\in\{1,2,R\}$, where $f_t$ is the number of data streams that needs to be decoded at the corresponding receiver. Thus, the received signal at $R_{j}$, in two time periods, can be written as
\begin{equation} 
\label{y_j}
\mathbf{y}_{j}^{[k]} =\underbrace{\sqrt{\frac{P_{j}}{d_{j,j}^{\tau_{j,j}}}} \mathbf{U}^{[k]H}_{j} \mathbf{H}^{[k]}_{j,j}\mathbf{V}^{[k]}_{j}\mathbf{s}_{j}}_{\text{desired signal}} + \underbrace{\mathbf{A}^{[k]}}_{\text{interference from SN}} 
+ \underbrace{ \sqrt{\frac{P_{i}}{d_{j,i}^{\tau_{j,i}}}} \mathbf{U}^{[k]H}_{j}\mathbf{H}^{[k]}_{j,i}\mathbf{V}^{[k]}_{i}\mathbf{s}_{i}}_{\text{interference from PN, \( i\neq j \)}}+ {\tilde{\mathbf{n}}}^{[k]}_{j},~k\in\{1,2\},
\end{equation}
where the effective noise term  $\tilde{\mathbf{n}}^{[k]}_{j}=\mathbf{U}^{[k]\,H}_{j}\mathbf{n}^{[k]}_{j}$ is a zero-mean additive white Gaussian noise (AWGN) vector, with $\mathbb{E}\{\tilde{\mathbf{n}}^{[k]}_{j}\tilde{\mathbf{n}}^{[k]H}_{j}\}=\sigma^2_{\tilde{n}_j}\mathbf{I}$, where $\mathbb{E}\{\cdot\}$ denotes an expectation operator. Moreover, we have $\mathbb{E}\{\mathbf{s}_l\mathbf{s}_l^H\}=\mathbf{I}$, with $l\in\mathcal{A}$, since $\mathbf{s}_l$ is assumed to be a vector consisting of symbols generated as i.i.d. Gaussian inputs. Finally, interference from SN to $R_{j}$ can be determined as
\begin{equation}
\mathbf{A}^{[k]} = \begin{cases}
 \sqrt{\frac{P_{S}}{d_{j,S}^{\tau_{j,S}}}} \mathbf{U}^{[k]H}_{j}\mathbf{H}_{j,S}\mathbf{V}_S\mathbf{s}_S,~\textrm{if}~k=1,\\
 \sqrt{\frac{P_{R}}{d_{j,R}^{\tau_{j,R}}}} \mathbf{U}^{[k]H}_{j}\mathbf{H}_{j,R}\mathbf{V}_R\mathbf{s}_R,~\textrm{if}~k=2.
\end{cases}
\end{equation}

The received signal at $R$, within transmission period of $S$-$R$, can be written as
\begin{equation}
\mathbf{y}_{R} =  \underbrace{ \sqrt{\frac{P_{S}}{d_{R,S}^{\tau_{R,S}}}} \mathbf{U}_{R}^{H}\mathbf{H}_{R,S}\mathbf{V}_S\mathbf{{s}}_S}_{\text{desired signal}} + \underbrace{ \sqrt{\frac{P_{i}}{d_{R,i}^{\tau_{R,i}}}} \sum_{i=1}^{2}\mathbf{U}_{R}^{H}\mathbf{H}_{R,i}\mathbf{V}^{[1]}_{i}\mathbf{s}_{i}}_{\text{interference from PN}} + \tilde{\mathbf{n}}_{R},
\end{equation}
where $\tilde{\mathbf{n}}_{R} = \mathbf{U}_{R}^{H}\mathbf{n}_{R}$ is the effective noise after interference suppression beamforming at the relay. 

Then, $R$ decodes and forwards the desired signal $\mathbf{s}_S$ to  $D$ within $R-D$ transmission period. Thus, $D$ obtains the following signal
\begin{align}
\mathbf{y}_{D} = \sqrt{\frac{P_{R}}{d_{D,R}^{\tau_{D,R}}}} \mathbf{H}_{D,R}\mathbf{V}_{R}\mathbf{s}_R+\mathbf{n}_{D},
\end{align}
where $\mathbf{n}_{D}$ is the AWGN vector, with $\mathbb{E}\{\mathbf{{n}}_{D}\mathbf{{n}}^H_{D}\}=\sigma^2_{{D}}\mathbf{I}$.

The interference in receive nodes can be assumed to be completely canceled if the following conditions are satisfied for $R_{j}$ as \cite{galym3,galym4}
\begin{subequations}
	\label{Rj condition}
	\begin{align} 
	&\mathbf{U}^{[k]H}_{j}\mathbf{H}^{[k]}_{j,i}\mathbf{V}^{[k]}_{i} = \mathbf{0},~\forall i,j\in\{1,2\},~\forall i\not=j,\\
	&\mathbf{U}^{[k]H}_{j} \mathbf{J}^{[k]} = \mathbf{0}, ~\text{where}~ \mathbf{J}^{[k]} = \begin{cases}\mathbf{H}_{j,S} \mathbf{V}_{S},~\text{if}~k = 1,\\
	\mathbf{H}_{j,R} \mathbf{V}_{R},~\text{if}~k = 2,
	\end{cases}\\
	&\textrm{rank}\left(\mathbf{U}_{j}^{[k]H}\mathbf{H}^{[k]}_{j,j}\mathbf{V}^{[k]}_{j}\right) = f_j,~\forall j\in\{1,2\}, 
	\end{align}
\end{subequations}
and for $R$ as
\begin{subequations}
	\label{R condition}
	\begin{align}
	&\mathbf{U}_{R}^{H}\mathbf{H}_{R,i}\mathbf{V}^{[1]}_{i} = \mathbf{0},~\forall i\in\{1,2\},\\
	&\textrm{rank}\left(\mathbf{U}_{R}^{H}\mathbf{H}_{R,S}\mathbf{V}_{S}\right) = f_S.    
	\end{align}
\end{subequations}

\subsection{Beamforming Design}
\label{sec:Transmit Beamforming Design}
If the space of the desired signal is linearly independent from that of the interference signal, then the desired signal can be easily decoded from the received one.  Hence, the design of precoding matrices should be in such a way that interference in all receivers need to span to one another. Thus, in the first time period, the interference at $R_1$, $R_2$ and $R$ can be spanned as $\text{span}\left(\mathbf{H}_{1,2}^{[1]}\mathbf{V}_{2}^{[1]}\right) =  \text{span}\left(\mathbf{H}_{1,S}\mathbf{V}_{S}\right)$, $\text{span}\left(\mathbf{H}_{2,1}^{[1]}\mathbf{V}_{1}^{[1]}\right) = \text{span}\left(\mathbf{H}_{2,S}\mathbf{V}_{S}\right)$ and $\text{span}\left(\mathbf{H}_{R,1}\mathbf{V}_{1}^{[1]}\right) =  \text{span}\left(\mathbf{H}_{R,2}\mathbf{V}_{2}^{[1]}\right)$, respectively, where $\text{span}(\mathbf{X})$ is the vector space spanned by the column vectors of $\mathbf{X}$. After spanning all interference, the precoding matrices $\mathbf{V}_{1}^{[1]}$, $\mathbf{V}_{2}^{[1]}$ and $\mathbf{V}_{S}$ can be obtained as \cite{Sung}
\begin{subequations}\label{V matirces_1}
	\begin{align}
	\mathbf{V}_{2}^{[1]} &= (\mathbf{H}_{R,2})^{-1}\mathbf{H}_{R,1}\mathbf{V}_{1}^{[1]},\\
	\mathbf{V}_{S} &= (\mathbf{H}_{2,S})^{-1}\mathbf{H}_{2,1}^{[1]}\mathbf{V}_{1}^{[1]},
	\end{align}
\end{subequations}
where $\mathbf{V}_{1}^{[1]}$ is derived using $\mathbf{V}_{1}^{[1]} = \text{eig}\left(\mathbf{Z}\right)$, with $\mathbf{Z} = \left(\mathbf{H}_{R,1}\right)^{-1}\mathbf{H}_{R,2}\left(\mathbf{H}_{1,2}^{[1]}\right)^{-1}\mathbf{H}_{1,S}(\mathbf{H}_{2,S})^{-1}\mathbf{H}_{2,1}^{[1]}$ and $\text{eig}(\mathbf{X})$ are the eigenvectors of $\mathbf{X}$.  

Interference suppression matrices ${\mathbf{U}_{j}^{[k]}}$ during two time slots, need to be orthogonalized to the interference at $R_{j}$ to meet conditions in \eqref{Rj condition}. Similarly, $\mathbf{U}_{R}$ needs to be orthogonalized to the interference at $R$ in $S$-$R$ transmission period. Derivations of those matrices can be written as
\begin{subequations}
	\label{U matirces_1}
	\begin{align}
	\mathbf{U}_{j}^{[k]} &= \text{null}\left( \left[\mathbf{H}_{j,i}^{[k]}\mathbf{V}_{i}^{[k]}\right]^H\right),~j\not=i,\\
	\mathbf{U}_{R} &= \text{null}\left( \left[\mathbf{H}_{R,1}\mathbf{V}_{1}^{[1]}\right]^H\right).	
	\end{align}
\end{subequations}

In the $2^{nd}$ time period, $S$ stays silent, while $R$ establishes its own communication. The design of precoding and interference suppression matrices for this time period can be done by following the same step in \eqref{V matirces_1}-\eqref{U matirces_1}.

\subsection{Imperfect CSI}
The assumption of perfect CSI in wireless networks is highly idealistic due to channel estimation error. Thus, the following model can be deployed for an imperfect CSI estimation \cite{galym2}
\begin{align}
\hat{\mathbf{H}}=\mathbf{H}+\mathbf{E},
\end{align}     
where $\hat{\mathbf{H}}$ is the observed mismatched channel, $\mathbf{H}\sim \mathcal{CN}(0,\mathbf{I})$ represents the real channel matrix and $\mathbf{E}$ is the error matrix which represents an inaccuracy degree in the estimated CSI. It is also assumed that $\mathbf{E}$ is independent of $\mathbf{H}$. Considering the  signal-to-noise ratio (SNR), $\theta$, $\mathbf{E}$ is described as 
\begin{align}
\mathbf{E}\sim \mathcal{CN}(0,\lambda\mathbf{I}) {\text{~with~}} \lambda=\psi{\theta}^{-\kappa},
\end{align} 
where $\lambda$ is an error variance, $\kappa\geq 0$ and $\psi>0$ determine various CSI scenarios. Finally, the real channel matrix, conditioning on ${\hat{\mathbf{H}}}$, \cite{Kay}, can be described as
\begin{align}
\label{Imperfect CSI}
\mathbf{H}=\frac{1}{1+\lambda} \hat{\mathbf{H}}+\tilde{\mathbf{H}},
\end{align}
where $\tilde{\mathbf{H}}\sim\mathcal{CN}(0, \frac{\lambda}{1+\lambda} \mathbf{I})$ is independent of $\hat{\mathbf{H}}$.
\begin{figure}[!h]
	\centering
	\includegraphics[width=0.6\columnwidth]{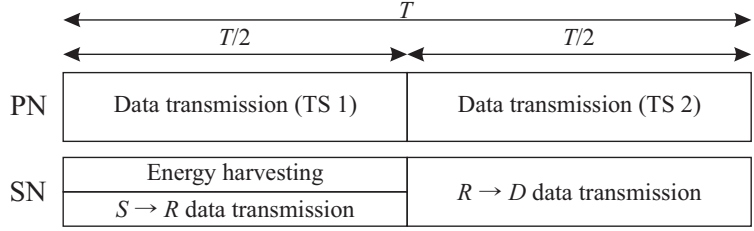}
	\caption{Time frame structure of PSR.}
	\label{subfig:PSR}
\end{figure}

\section{Power-Splitting Relaying}
\label{sec:PSR}
The PSR for SWIPT is shown in Fig. \ref{subfig:PSR}, where the total  time is split into two equal portions, one for the $S$-$R$ and the rest for $R$-$D$ data transmissions \cite{Khaled}. Within the $1^{st}$ time fraction, an energy portion of $\rho$, with  $0<\rho<1$, at $R$ is allocated for EH, while the remaining power of $(1 - \rho)$ is conveyed to data transmission purpose.

Hence, the $R$ obtains the following signal for EH 
\begin{equation}
\label{y_r_eh_psr}
\mathbf{y}_R^{EH}=\sqrt{\frac{\rho P_S}{d^{\tau_{R,S}}_{R,S}}}\mathbf{H}_{R,S}\mathbf{V}_{S}\mathbf{s}_S + \sum_{i=1}^{2}\sqrt{\frac{\rho P_i}{d^{\tau_{R,i}}_{R,i}}}\mathbf{H}_{R,i}\mathbf{V}^{[1]}_{i}\mathbf{s}_i+\sqrt{\rho}\mathbf{n}_{R}.
\end{equation}

The power harvested from the noise is insignificant which can be neglected. Thus, the instantaneous harvested energy at $R$ can be derived from \eqref{y_r_eh_psr} as \cite{Nasir}
\begin{equation}
\label{E_h}
P_R = \eta\rho\left( \frac{P_S}{d^{\tau_{R,S}}_{R,S}}\left| \left|\mathbf{H}_{R,S}\mathbf{V}_S\right| \right| ^2 + \sum_{i=1}^{2}\frac{P_i}{d^{\tau_{R,i}}_{R,i}}\left| \left| \mathbf{H}_{R,i}\mathbf{V}^{[1]}_i\right| \right|^2 \right), 
\end{equation}
where $||\cdot||$ denotes the Euclidean norm. 
Then, by using \eqref{Imperfect CSI}, the received information signal with power $(1-\rho)$ at $R$ can be derived as  \eqref{y_IT} at the top of the next page.

\begin{figure*}[!t]
	\small\begin{align}
	\label{y_IT}
	\mathbf{y}^{IT}_R =& \sqrt{1-\rho} \mathbf{U}^H_R\left( \sqrt{\frac{P_S}{d^{\tau_{R,S}}_{R,S}}} \left( \frac{1}{1+\lambda}\hat{\mathbf{H}}_{R,S} + \tilde{\mathbf{H}}_{R,S}\right) \mathbf{V}_{S}\mathbf{s}_S + \sum_{i=1}^{2} \sqrt{\frac{P_i}{d^{\tau_{R,i}}_{R,i}}}  \left( \frac{1}{1+\lambda}\hat{\mathbf{H}}_{R,i}+\tilde{\mathbf{H}}_{R,i} \right) \mathbf{V}^{[1]}_{i}{\mathbf{s}_i} + \mathbf{n}_{R} \right)
	\end{align}
	\hrulefill
\end{figure*}

The corresponding signal-to-interference-noise ratio (SINR) for $R$ from \eqref{y_IT} is derived by following
\begin{align}
\label{gamma_r_ps}
\gamma_R=\frac{\frac{P_S(1-\rho)}{d^{\tau_{R,S}}_{R,S} (1+\lambda)^2}|| \mathbf{U}_R^H\hat{\mathbf{H}}_{R,S}\mathbf{V}_S|| ^2}{\frac{P_S (1-\rho)}{d^{\tau_{R,S}}_{R,S}} || \mathbf{U}_R^H\tilde{\mathbf{H}}_{R,S}\mathbf{V}_S||^2 + I_{PN} + \sigma^2_{\tilde{n}_R}},
\end{align}
where $I_{PN} = \frac{P_i (1-\rho)}{d^{\tau_{R,i}}_{R,i}} \sum_{i=1}^{2} || \mathbf{U}_R^H\tilde{\mathbf{H}}_{R,i}\mathbf{V}^{[1]}_i||^2$ defines the interference from primary transmitters.

Then, the received signal at $D$ can be written as
\begin{align}
\label{y_d}
\mathbf{y}_{D}=&\sqrt{\frac{P_R}{d^{\tau_{D,R}}_{D,R}}}\left( \frac{1}{1+\lambda}\hat{\mathbf{H}}_{D,R}+{\tilde{\mathbf{H}}}_{D,R}\right) \mathbf{V}_{R}\mathbf{{s}}_R+\mathbf{n}_{D}.
\end{align} 
and SINR from \eqref{y_d} can be derived as
\begin{align}
\label{gamma_d_ps}
\gamma_D = \frac{\frac{P_R}{d^{\tau_{D,R}}_{D,R} (1+\lambda)^2 } || {\hat{\mathbf{H}}}_{D,R}\mathbf{V}_R||^2}{ \frac{P_R}{d^{\tau_{D,R}}_{D,R}} || \tilde{\mathbf{H}}_{D,R}\mathbf{V}_R||^2+ \sigma^2_{D} } ,
\end{align}
where $\sigma^2_{D}$ is the noise power. 

Also, the received SINR for $R_{j}$ is shown as
\begin{align}
\label{gamma_j_ps}
\gamma^{[k]}_j=\frac{\frac{P_j}{d^{\tau_{j,j}}_{j,j} (1+\lambda)^2}|| {\mathbf{U}^{[k]H}_j}\hat{\mathbf{H}}^{[k]}_{j,j}\mathbf{V}^{[k]}_j||^2}{B^{[k]} + C^{[k]} + {\sigma_{\tilde{n}_{j}}^2}^{[k]}},
\end{align}
where the intra-network interference of the PN due to the CSI mismatch is given by $B^{[k]} = \frac{P_j}{d^{\tau_{j,j}}_{j,j}} || {\mathbf{U}^{[k]H}_j}\tilde{\mathbf{H}}^{[k]}_{j,j}\mathbf{V}^{[k]}_j ||^2 + 
\frac{P_i}{d^{\tau_{j,i}}_{j,i}} || {\mathbf{U}^{[k]H}_j}\tilde{\mathbf{H}}^{[k]}_{j,i}\mathbf{V}^{[k]}_i ||^2_{i\not=j}$ while the inter-network interference from the SN is expressed by
\begin{equation}
C^{[k]} = \begin{cases}
{\frac{P_S}{d^{\tau_{j,S}}_{j,S}} || {\mathbf{U}^{[k]H}_j}\tilde{\mathbf{H}}_{j,S}\mathbf{V}_S|| ^2},~\textrm{if}~k=1,\\
{\frac{P_R}{d^{\tau_{j,R}}_{j,R}} || {\mathbf{U}^{[k]H}_j}\tilde{\mathbf{H}}_{j,R}\mathbf{V}_R||^2},~\textrm{if}~k=2.
\end{cases}
\end{equation}

The BER of symbol $\mathbf{s}_m$ for binary phase shift keying (BPSK) can be derived as \cite{Ohno}
\begin{align}
\text{BER}_m = Q(\sqrt{\gamma_m}),~m\in\{1,2,R,D\},
\end{align}
where $Q(x) = 1/ \sqrt{2\pi} \int_{x}^{\infty} \text{exp}(-t^2/2) dt$ is the Gaussian-$Q$ function.

\section{Simulation Results}
\label{sec:numerical}
This section presents the simulation results for the proposed system model in Rayleigh fading channels with BPSK modulation. The system parameters are as follows: $d_{m,n}=3$ and $\tau_{m,n}=2.7~\forall m\in\{1,2,R,D\},~\forall n\in\{1,2,S,R\}$ and equal transmit power at $T_i$ and $S$. The calculated optimal values of $\rho=0.75$ with $\eta=0.8$ in \cite{sultan2} is considered. Furthermore, the following values of $(\kappa, \psi)$ such as $(0,0.001),~(0,0.05),~(0.75,10),~(1,10)$ and $(1.5,15)$ are used to investigate the impact of CSI mismatch.

\begin{figure*}[!h]
	\centering
	\subfloat[BER performance for perfect CSI and SNR-independent CSI mismatch $\left( (0,~0.001)~\text{and}~(0,~0.005)\right) $.]{
		\label{subfig:kappa zero}
		\includegraphics[width=0.45\columnwidth]{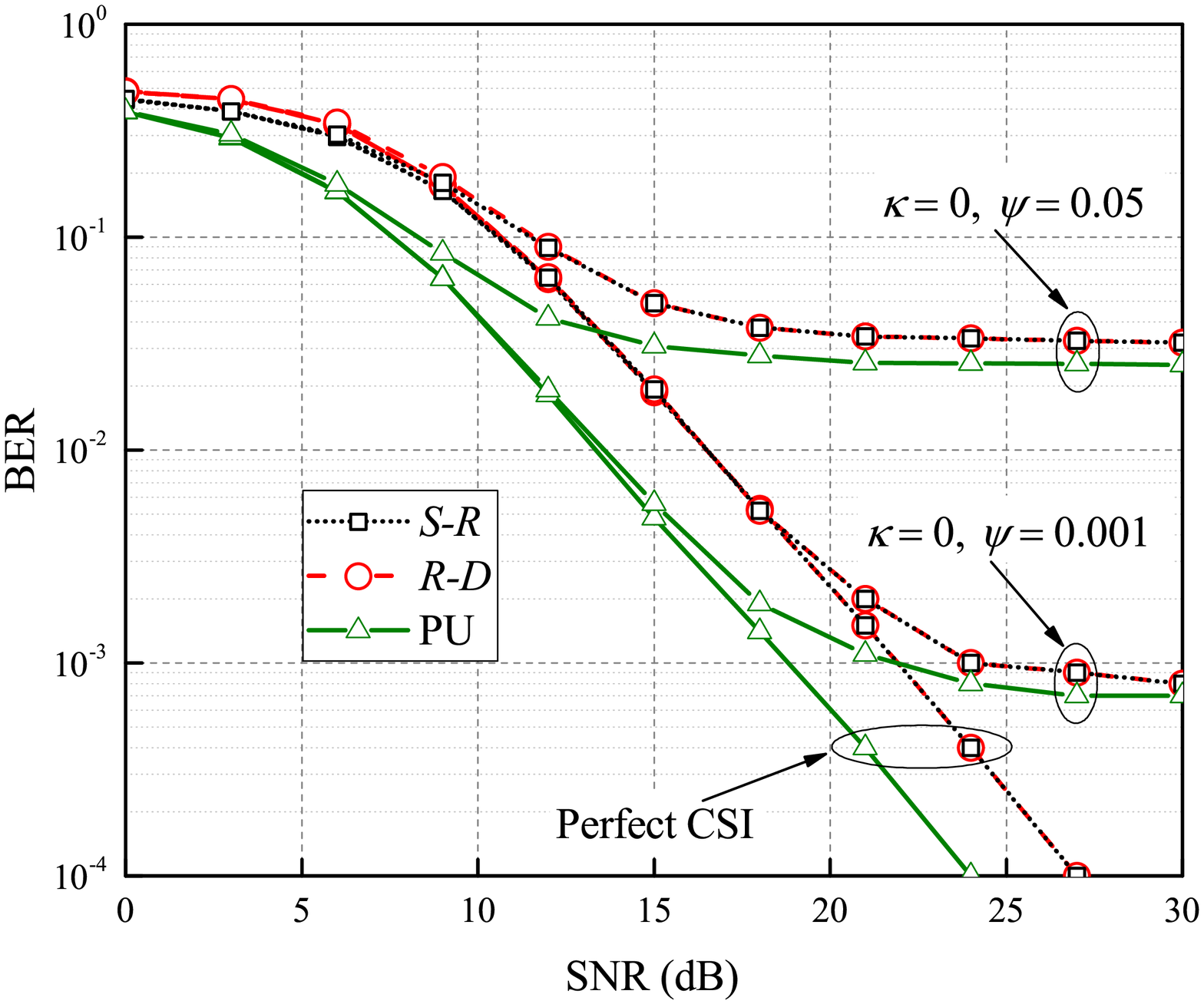}}~~~~
	\subfloat[BER performance for SNR-dependent CSI mismatch $\left( (0.75,~10),~(1,~10)~\text{and}~(1.5,~15)\right) $.]{
		\label{subfig:kappa diff}
		\includegraphics[width=0.45\columnwidth]{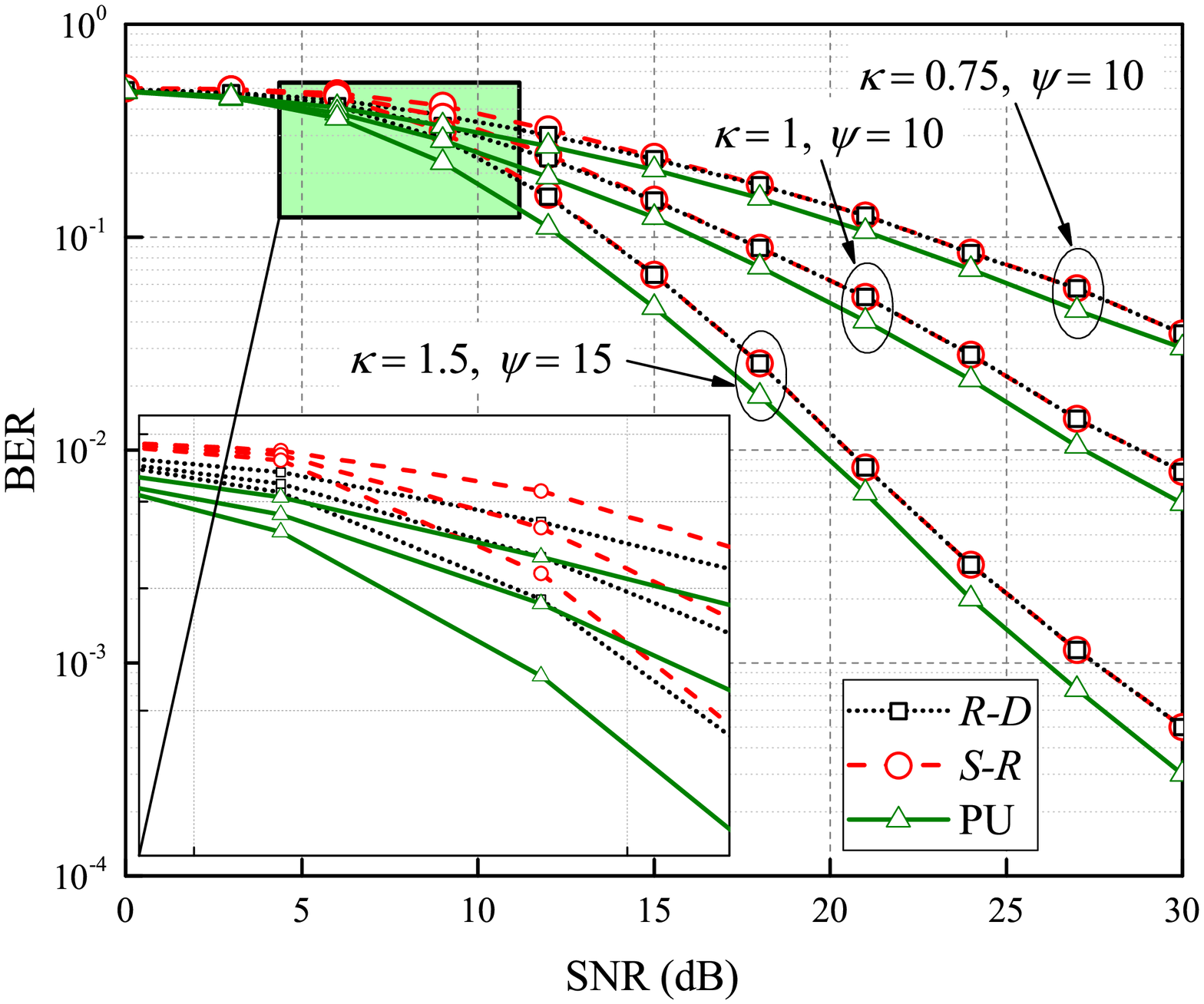}}
	\caption{BER vs. SNR of the primary user, the relay and the destination node operating in the PSR protocol under different CSI scenarios.}
	\label{results1}
\end{figure*}

Fig. \ref{results1} shows how imperfect CSI parameters impact on the BER performance of the PU and SUs. For the case of SNR-independent CSI mismatch (when $\kappa=0$, Fig. \ref{subfig:kappa zero}), the BER degrades as $\psi$ increases because the channel error variance does not depend on the SNR and the BER curves saturate after some SNR values, e.g. at 15 dB and 21 dB for 0.05 and 0.001, respectively. Furthermore, it is worth to note that the BER performance is not affected by $\psi$ in the low SNR region, i.e. $\psi$ starts playing a role at 3 dB and 6 dB for the BER of PU and SUs, respectively. This can be explained by the fact that small values of $\psi$ do not increase much the error rate of the received signal at low SNR. 
 
In general, BER performance of PU outperforms those of $R$ and $D$ because of the power portion $1-\rho$ devoted for data transmission at $R$. The less the value of the allocated power for data transmission is, the higher the probability of the incorrect data detection becomes. To compare the performance of $R$ and $D$ at low SNR, $R$ performs  better than $D$ because $R$ transmits its data with a certain number of errors which in turn affects the error rate of $D$. However, the BER performances of SUs match after 10 dB due to the fact that $R$ harvests more energy at high SNR and $D$ consequently receives a strong signal to detect. When $\kappa\neq0$, the channel error variance becomes SNR-dependent (see Fig. \ref{subfig:kappa diff}), which implies no saturation of the BER performance. An increase of $\kappa$ leads to the BER improvement. At 30 dB, the BER performance of SUs obtains 0.0353, 0.0079 and 0.0005 for ($0.75, 10$), ($1, 10$) and ($1.5, 15$), respectively.
  
\begin{figure}[!t]
\centering
\includegraphics[width=0.6\columnwidth]{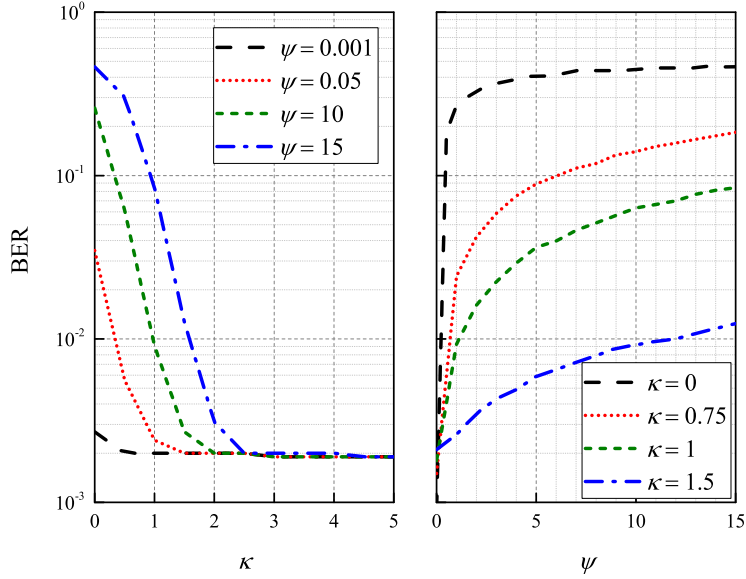}
\caption{BER vs. the CSI mismatch parameters $\kappa$ and $\psi$ of SU at 20 dB.}
	\label{kappa_psi}
\end{figure}

A more deeper analysis on the impact of $\kappa$ and $\psi$ on the BER performance can be obtained from Fig. \ref{kappa_psi}, where the BER performance of SUs is built up as a function of different values of $\kappa$ and $\psi$ at 20 dB. It can be noticed that the BER performance improves as $\kappa$ increases, while an increase of $\psi$ results in the BER degradation. In the first subfigure, the BER curves for different values of $\psi$ approach 0.0019 at certain $\kappa$ values. 
Meanwhile, in the second subfigure, the BER performance for different $\kappa$ degrades as $\psi$ increases. It is observed that small values of $\kappa$ correspond to more abrupt BER degradation, and vise versa.

\section{Conclusion}
\label{sec:Conclusion}
In this paper, we analyzed the BER performance of EH-based DF CRN with PS relaying protocols and embedded IA technique. The five special scenarios of the imperfect CSI given by $(0,0.001),~(0,0.05),~(0.75,10),~(1,10)$ and $(1.5,15)$ were studied to analyze the impact of the CSI quality on the BER performance of PU and SUs. The presented results with $\rho=0.75$ showed that the BER of PU outperforms those of SUs in perfect and imperfect SCI cases. Moreover, the BER curve degraded as $\psi$ increased while rise of $\kappa$ leaded to the BER improvement.    

\ifCLASSOPTIONcaptionsoff
\newpage
\fi


\begin{thebibliography}{1}
	\bibitem{Goldsmith}   
	A.~Goldsmith, S.~Jafar, L.~Maric and S.~Srinivasa, ``Breaking spectrum gridlock with cognitive radios: an information theoretic perspective,'' \emph{IEEE Proceedings}, vol. 97, no. 5, pp. 894--914, May 2009.
	\bibitem{Arzykulov}           
	S.~Arzykulov, G.~Nauryzbayev and T.~A.~Tsiftsis, ``Underlay Cognitive Relaying System Over $\alpha$ - $\mu$ Fading Channels,'' \emph{IEEE Commun. Lett.,} vol. 21, no. 1, pp. 216--219, Jan. 2017
	\bibitem{galym1}   
	G.~Nauryzbayev and E.~Alsusa, ``Enhanced Multiplexing Gain Using Interference Alignment Cancellation in Multi-cell MIMO Networks,'' \emph{IEEE Trans. Inf. Theory}, vol. 62, no. 1, pp. 357--369, Jan. 2016.
	\bibitem{galym2}   
	G.~Nauryzbayev and E.~Alsusa, ``Interference Alignment Cancellation in Compounded MIMO Broadcast Channels with General Message Sets,'' \emph{IEEE Trans. Commun.}, vol. 63, no. 10, pp. 3702--3712, Oct. 2015.
	\bibitem{Amir}
	M.~Amir, A.~El-Keyi and M.~Nafie, ``Constrained Interference Alignment and the Spatial Degrees of Freedom of MIMO Cognitive Networks,'' \emph{IEEE Trans. Inf. Theory,} vol. 57, no. 5, pp. 2994--3004, May 2011.
	\bibitem{Tang}        
	J.~Tang, S.~Lambotharan and S.~Pomeroy,  ``Interference cancellation and alignment techniques for multiple-input and multiple-output cognitive relay networks,'' \emph{IET Signal Process.,} vol. 7, no. 3, pp. 188--200, May 2013.
	\bibitem{Nasir} 
	A.~Nasir, X.~Zhou, S.~Durrani, and R.~Kennedy, ``Relaying protocols for wireless energy harvesting and information processing,'' \emph{IEEE Trans. Wireless Commun.,} vol. 12, no. 7, pp. 3622--3636, Jul. 2013.
	\bibitem{GN}
	G.~Nauryzbayev, K.~M.~Rabie, M.~Abdallah and B.~Adebisi, ``Ergodic Capacity Analysis of Wireless Powered AF Relaying Systems over $\alpha$ - $\mu$ Fading Channels,'' in \emph{Proc. IEEE Global Commun. Conf. (GLOBECOM),} Singapore, pp. 1--6, Dec. 2017.
	\bibitem{Zhao1}
	N.~Zhao, F.~R.~Yu and V.~C.~M.~Leung, ``Wireless energy harvesting in interference alignment networks,'' \emph{IEEE  Commun. Mag.}, vol. 53, no. 6, pp. 72-78, June 2015.
	\bibitem{Park}        
	S.~Park, H.~Kim and D.~Hong, ``Cognitive radio networks with energy harvesting,'' \emph{IEEE Trans. Wireless Commun.,} vol. 12, no. 3, pp. 1386--1397, Mar. 2013.    
	\bibitem{Zheng}         
	G.~Zheng, Z.~Ho, E.~A.~Jorswieck and B.~Ottersten, ``Information and energy cooperation in cognitive radio networks,'' \emph{IEEE Trans. Signal Process.,} vol. 62, no. 9, pp. 2290--2303, Sept. 2014.    
	\bibitem{Wang}         
	F.~Wang and X.~Zhang, ``Resource Allocation for Multiuser Cooperative Overlay Cognitive Radio Networks with RF Energy Harvesting Capability,'' \emph{IEEE Global Commun. Conf. (GLOBECOM),} Washington, DC, pp. 1--6, 2016.
	\bibitem{galym3}
	G. Nauryzbayev and E. Alsusa, ``Identifying the Maximum DoF Region in the Three-cell Compounded MIMO Network,'' \emph{IEEE WCNC,} pp. 1--5, Doha, Qatar, Apr. 2016.
	\bibitem{galym4}
	G. Nauryzbayev, E. Alsusa, and J. Tang, ``An Alignment Based Interference Cancellation Scheme for Multi-cell MIMO Networks,'' \emph{IEEE VTC,} pp. 1--5, Glasgow, UK, May 2015.
	\bibitem{Sung}
	H.~Sung, S.~H.~Park, K.~J.~Lee and I.~Lee, ``Linear precoder designs for K-user interference channels,'' \emph{IEEE Trans. Wireless Commun.}, vol. 9, no. 1, pp. 291--301, Jan. 2010. 
 	\bibitem{Kay}         
	S.~Kay, ``Fundamentals of statistical signal processing: Estimation theory'' in Englewood Cliffs NJ: Prentice-Hall 1993.
  	\bibitem{Ohno}
	S.~Ohno and K.~A.~D.~Teo,  ``Universal BER Performance Ordering of MIMO Systems over Flat Channels,'' \emph{IEEE Trans. Wireless Commun.,} vol. 6, no. 10, pp. 3678-3687, Oct. 2007.
    \bibitem{sultan2}
    S.~Arzykulov, G.~Nauryzbayev, T.~A.~Tsiftsis and M.~Abdallah, ``On the Capacity of Wireless Powered Cognitive Relay Network with Interference Alignment,'' in \emph{Proc. IEEE Global Commun. Conf. (GLOBECOM),} Singapore, pp. 1--6, Dec. 2017.

\end{thebibliography}
\end{document}